\documentclass{ws-procs9x6}

\def\etal {{\it et al.}}

\def\kop{\tilde{\kappa}_{o+}}

\def\lsim{\mathrel{\rlap{\lower3pt\hbox{$\sim$}}
   \raise2pt\hbox{$<$}}}
\def\gsim{\mathrel{\rlap{\lower3pt\hbox{$\sim$}}
   \raise2pt\hbox{$>$}}}
\def\sqr#1#2{{\vcenter{\vbox{\hrule height.#2pt
        \hbox{\vrule width.#2pt height#1pt \kern#1pt
        \vrule width.#2pt}
        \hrule height.#2pt}}}}

\def\etal{{\it et al.}}

\newcommand{\beq}[1]{\begin{equation}\label{#1}}
\newcommand{\eeq}{\end{equation}}
\newcommand{\bea}[1]{\begin{eqnarray}\label{#1}}
\newcommand{\eea}{\end{eqnarray}}
\newcommand{\ba}{\begin{array}}
\newcommand{\ea}{\end{array}}

\newcommand{\rf}[1]{(\ref{#1})}

\begin{document}

\title{Search for light-speed anisotropies 
using Compton scattering of high-energy electrons}

\author{D. Rebreyend on behalf of the experimental team\footnote{J.-P.~Bocquet, D.~Moricciani, 
V.~Bellini, M.~Beretta, L.~Casano, A.~D'Angelo, R.~Di Salvo, A.~Fantini, D.~Franco,
G.~Gervino, F.~Ghio, G.~Giardina, B.~Girolami, A.~Giusa, V.G.~Gurzadyan, A.~Kashin, S.~Knyazyan, A.~Lapik,
R.~Lehnert, P.~Levi Sandri, A.~Lleres, F.~Mammoliti, G.~Mandaglio, M.~Manganaro, A.~Margarian, S.~Mehrabyan,
R.~Messi, V.~Nedorezov, C.~Perrin, C.~Randieri, N.~Rudnev, G.~Russo, C.~Schaerf,
M.-L.~Sperduto, M.-C.~Sutera, A.~Turinge, and V.~Vegna}}

\address{Laboratoire de Physique Subatomique et de Cosmologie, UJF Grenoble 1, CNRS/IN2P3, INPG, 
Grenoble, France\\
E-mail: rebreyend@lpsc.in2p3.fr}

\begin{abstract}
Based on the high sensitivity of Compton scattering off ultra relativistic electrons, the possibility of anisotropies in the speed of light 
is investigated. The result discussed in this contribution is based on the $\gamma$-ray beam of the ESRF's GRAAL facility (Grenoble, France) and  the 
search for sidereal variations in the energy of the Compton-edge photons. The absence of oscillations
yields the two-sided limit of $1.6 \times 10^{-14}$  at $95\,\%$ confidence level
on a combination of photon and electron coefficients of the minimal Standard Model Extension (mSME).
This new constraint provides an improvement over previous bounds 
by one order of magnitude. 
\end{abstract}

\bigskip

\bodymatter

Experimental searches for anisotropies in $c$ and, more generally, for Lorentz violating (LV) processes
are currently motivated by
theoretical studies in the context of quantum gravity. 
Recent approaches to Planck-scale physics 
can indeed accommodate minuscule violations of Lorentz symmetry~\cite{lotsoftheory}.

The present result is based on a laboratory experiment using only
photons and electrons in an environment 
where gravity is negligible. 
Lorentz violation can then be described by
the single-flavor QED limit of the flat-spacetime mSME~\cite{sme,collider}.
In this framework, photons have a modified dispersion relation:
\beq{ModDR}
\omega=(1-\vec{\kappa}\cdot\hat{\lambda})\,\lambda+{\mathcal O}(\kappa^2)\;.
\eeq
\noindent Here, $\lambda^\mu=(\omega,\lambda\hat{\lambda})$ denotes the photon 4-momentum
and $\hat{\lambda}$ is a unit 3-vector. 
The space-time constant mSME $\vec{\kappa}$ vector specifies a preferred direction in the Universe
which violates Lorentz symmetry, and can be interpreted 
as generating a direction-dependent refractive index of the vacuum 
$n(\hat{\lambda})\simeq1+\vec{\kappa}\cdot\hat{\lambda}$. 

The basic experimental idea is that
in a terrestrial laboratory 
the photon 3-momentum in a Compton-scattering process 
changes direction due to the Earth's rotation.  
The photons are thus affected by the anisotropies in Eq.~\rf{ModDR}
leading to sidereal effects in the kinematics of the process. 

The experimental set-up at GRAAL
involves counter-propagating incoming electrons and photons 
with 3-momenta $\vec{p}=p\,\hat{p}$ and $\vec{\lambda}=-\lambda\,\hat{p}$, 
respectively. The conventional Compton edge (CE) then occurs
for outgoing photons that are backscattered at $180^\circ$,
so that the kinematics is essentially one dimensional
along the beam direction $\hat{p}$.
Energy conservation for this process reads
\beq{Econs}
E(p)+(1+\vec{\kappa}\cdot\hat{p})\,\lambda
=E(p-\lambda-\lambda')+(1-\vec{\kappa}\cdot\hat{p})\,\lambda'\;,
\eeq
where $\vec{\lambda}'=\lambda'\,\hat{p}$ is the 3-momentum
of the CE photon, 
and 3-momentum conservation has been implemented.
At leading order, 
the physical solution of Eq.~\rf{Econs} is
\beq{modCE}
\lambda'\simeq\lambda_{\rm CE}
\left[\,
1 + \frac{2\,\gamma^2}{(1 + 4\, \gamma\, \lambda\, /\, m)^2}\,\vec{\kappa}\cdot\hat{p}
\,\right]\,.
\eeq
Here, 
$\lambda_{\rm CE} = \frac{4\, \gamma^2 \, \lambda }{1 + 4\, \gamma\, \lambda\, /\, m}$ 
denotes the conventional value of the CE energy. 
Given the actual experimental data of 
$m=511\,$keV, 
$p=6030\,$MeV, 
and $\lambda=3.5\,$eV, 
yields $\gamma \simeq p/m = 11800$ and $\lambda_{\rm CE}=1473\,$MeV.
The numerical value of the factor in front of $\vec{\kappa}\cdot\hat{p}$ 
is about $1.6\times10^8$.
It is this large  amplification factor (essentially given 
by $\gamma^2$) that 
yields the exceptional sensitivity of the CE 
to $\kop$.

Expressed in the Sun-centered inertial frame $(X,Y,Z)$~\cite{kr}
and taking into account GRAAL's latitude and beam direction,
eq.~\rf{modCE} becomes
\beq{sidereal}
\lambda'\simeq\tilde{\lambda}_{\rm CE}+
0.91\,\frac{2\,\gamma^2\,\lambda_{\rm CE}}{(1 + 4\,\gamma\,\lambda\, / \,m )^2}
\sqrt{\kappa_X^2+\kappa_Y ^2}
\,\sin\Omega t\,.
\eeq

\begin{figure}
\begin{center}
\includegraphics[width=0.9\hsize]{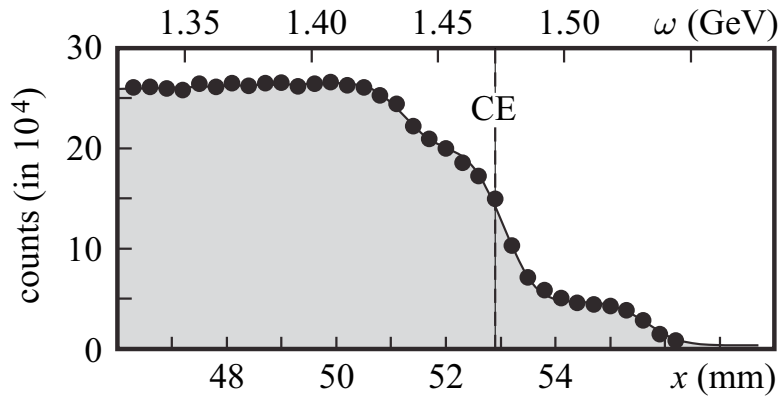}
\end{center}
\vskip-10pt
\caption{Si $\mu$-strip count spectrum near the CE 
and the fitting function (see text) vs.\ position $x$
and photon energy $\omega$. 
The three edges corresponding to the lines
$364$, $351$, and $333\,$nm are clearly visible. 
The CE position $x_{\rm CE}$ 
is the location of the central line 
and is measured with a typical accuracy of $3\,$$\mu$m.
\vspace{-5mm}}
\label{countrate}
\end{figure}

Incoming photons overlap with the ESRF beam 
over a $6.5\,$m long straight section.  
Due to their energy loss, 
scattered electrons are extracted from the main beam 
in the magnetic dipole 
following the straight section. 
Their position can then be accurately measured 
in the so-called tagging system 
%(Fig.~\ref{test}) 
located $50\,$cm after the exit of the dipole. 
This system is composed of a position-sensitive Si $\mu$-strip detector 
(128~strips of $300\,$$\mu$m pitch, $500\,$$\mu$m thick) 
associated to a set of fast plastic scintillators.
A typical Si $\mu$-strip count spectrum 
near the CE is shown in Fig.~\ref{countrate} 
for the multiline UV mode ($364$, $351$, $333\,$nm) of the laser used in this measurement. 
The fitting function, 
also plotted, 
is based on the sum of 3 error functions 
plus background 
and includes 6 free parameters. 
The CE position (location of the central line),
$x_{\rm CE}$, can be measured with an excellent resolution of $\sim 3\,\mu$m.

\begin{figure}
\begin{center}
\includegraphics[width=0.9\hsize]{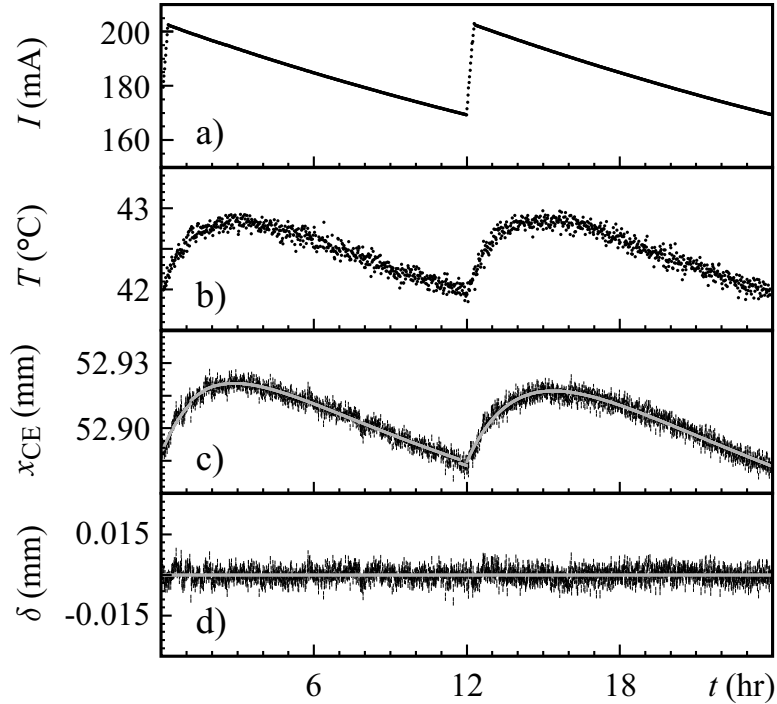}
\end{center}
\vskip-10pt
\caption{Time evolution over a day of 
a) ESRF beam intensity; 
b) tagging-box temperature; 
c) CE position and fitted curve; 
d) $\delta = x_{\rm CE} - x_{\rm fit}$. 
The error bars on position measurements are directly given by the CE fit.
\vspace{-5mm}}
\label{t_evol}
\end{figure}

A sample of the time series of the CE positions 
relative to the ESRF beam 
covering $24\,$h 
is displayed in Fig.~\ref{t_evol}c, 
along with the tagging-box temperature (Fig.~\ref{t_evol}b) 
and the ESRF beam intensity (Fig.~\ref{t_evol}a). 
The sharp steps present in Fig.~\ref{t_evol}a 
correspond to the twice-a-day refills of the ESRF ring. 
The similarity of the temperature and CE spectra
combined with their correlation with the ESRF beam intensity 
led us to interpret the continuous and slow drift of the CE positions 
as a result of the tagging-box dilation induced by the x-ray heat load.

To get rid of this trivial time dependence, raw data have been fitted
with the sum of two exponential whose time constants have been extracted
from the time evolution of the temperature data.
The corrected and final spectrum is 
obtained by subtraction of the fitted function from the raw data, (Fig.~\ref{t_evol}d). 

The usual equation for the deflection of charges in a magnetic field
together with momentum conservation in Compton scattering
determines the relation between the CE variations $\frac{\Delta x_{\rm CE}}{x_{\rm CE}}$ 
and a hypothetical CE photon 3-momentum oscillation $\frac{\Delta \lambda'}{\lambda'}$:
\beq{DLL}
\frac{\Delta x_{\rm CE}}{x_{\rm CE}} = 
\frac{p}{p - \lambda_{\rm CE}}\frac{\Delta \lambda'}{\lambda'}\,.
\eeq

\begin{figure}
\begin{center}
\includegraphics[width=0.9\linewidth]{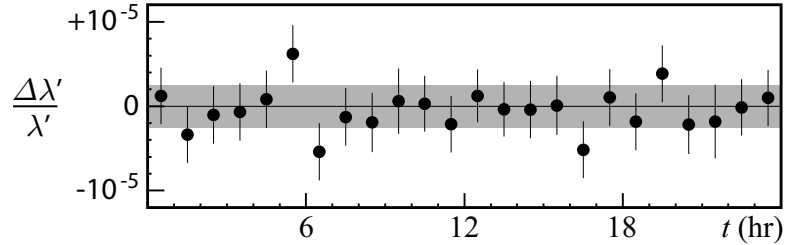}
\end{center}
\vskip-10pt
\caption{Full set of data folded modulo a sidereal day (24~bins). 
The error bars are purely statistical 
and agree with the dispersion of the data points ($\chi^2=1.04$ for the unbinned histogram).
The shaded area corresponds to the region of non-excluded signal amplitudes. \vspace{-5mm}}
\label{final}
\end{figure}

To search for a modulation, 14765 data points collected in about 1 week of data taking have been folded modulo a sidereal day 
(Fig.~\ref{final}). The error bars are purely statistical
and the histogram is in agreement with a null signal ($\chi^2=1.04$). 
To look for a harmonic oscillation ($A\sin(\Omega t + \phi)$), we have performed a statistical analysis
based on the
Bayesian approach. The resulting
upper bound is $A<2.5\times10^{-6}$ 
at $95\%$ CL.

We next consider effects that 
could conceal an actual sidereal signal.  
Besides a direct oscillation of the orbit,
the two quantities that 
may affect the result are
the dipole magnetic field   
and the momentum of the ESRF beam $p$. 
All these parameters are linked to the machine operation, 
and their stability follows directly from the accelerator performance. 
A detailed analysis of the ESRF database 
allows us to conclude that a sidereal oscillation of any of these parameters
cannot exceed a few parts in $10^{7}$ and is negligible.

We can now conclude that
our upper bound on a hypothetical sidereal oscillation of the CE energy is:
\beq{A_bound}
\Delta \lambda'/\lambda' < 2.5\times10^{-6}\quad (95\,\% \; {\rm CL})\,,
\eeq
yielding the competitive limit
$\sqrt{\kappa_X^2+\kappa_Y^2} < 1.6\times10^{-14}$ ($95\,\%$ CL)
with Eq.~\rf{sidereal} \cite{boc10}.
This limit improves previous bounds by a factor of ten and represents the first test of Special Relativity 
via a non-threshold kinematics effect in a particle collision \cite{thres}.

\end{document}